\documentclass[sigconf, nonacm]{acmart}

\AtBeginDocument{%
  }

\begin{document}

\title{The Value Semantics Layer: Enabling Machine-Understandable Databases for Agentic AI}

\author{Jalal Mahmud}
\affiliation{%
  \institution{Megagon Labs}
  \city{Mountain View}
  \state{California}
  \country{USA}
}
\email{jalal@megagon.ai}

\author{Eser Kandogan}
\affiliation{%
  \institution{Megagon Labs}
  \city{Mountain View}
  \state{California}
  \country{USA}
}

\email{eser@megagon.ai}



\begin{abstract}
As data systems increasingly support autonomous LLM-powered
agents alongside human analysts, they require richer
semantic infrastructure bridging analytical concepts and
analytical operations.
Existing metadata systems expose
schemas, catalogs, lineage, and governance information, but
provide limited support for representing operational
semantics. We propose the \emph{Value Semantics Layer}
(VSL), an architectural abstraction that externalizes
value semantics as persistent, planner-visible
infrastructure. VSL introduces \emph{Value Semantic Types}
(VSTs), reusable operational semantic assertions capturing
distributional, relational, process, quantitative, and
policy-grounded semantics. Unlike descriptive metadata,
VSTs enable planners to validate analytical operations,
select appropriate operators, and guide analytical
workflows. We present the motivation, formalization,
architecture, and prototype realizations of VSL, and
outline a research agenda toward AI-native semantic
infrastructure for analytical systems.
\end{abstract}



\maketitle

\section{Introduction}

Analytical workflows were traditionally authored by domain
experts, leaving much of the operational interpretation of data
values implicit in analyst expertise rather than represented in the
data system. As natural language becomes the interface for human
analysts and autonomous agents alike, this responsibility shifts to
computational systems, which must interpret analytical concepts
before generating executable queries or workflows.

\begin{figure*}[t]
\centering
\includegraphics[width=\linewidth]{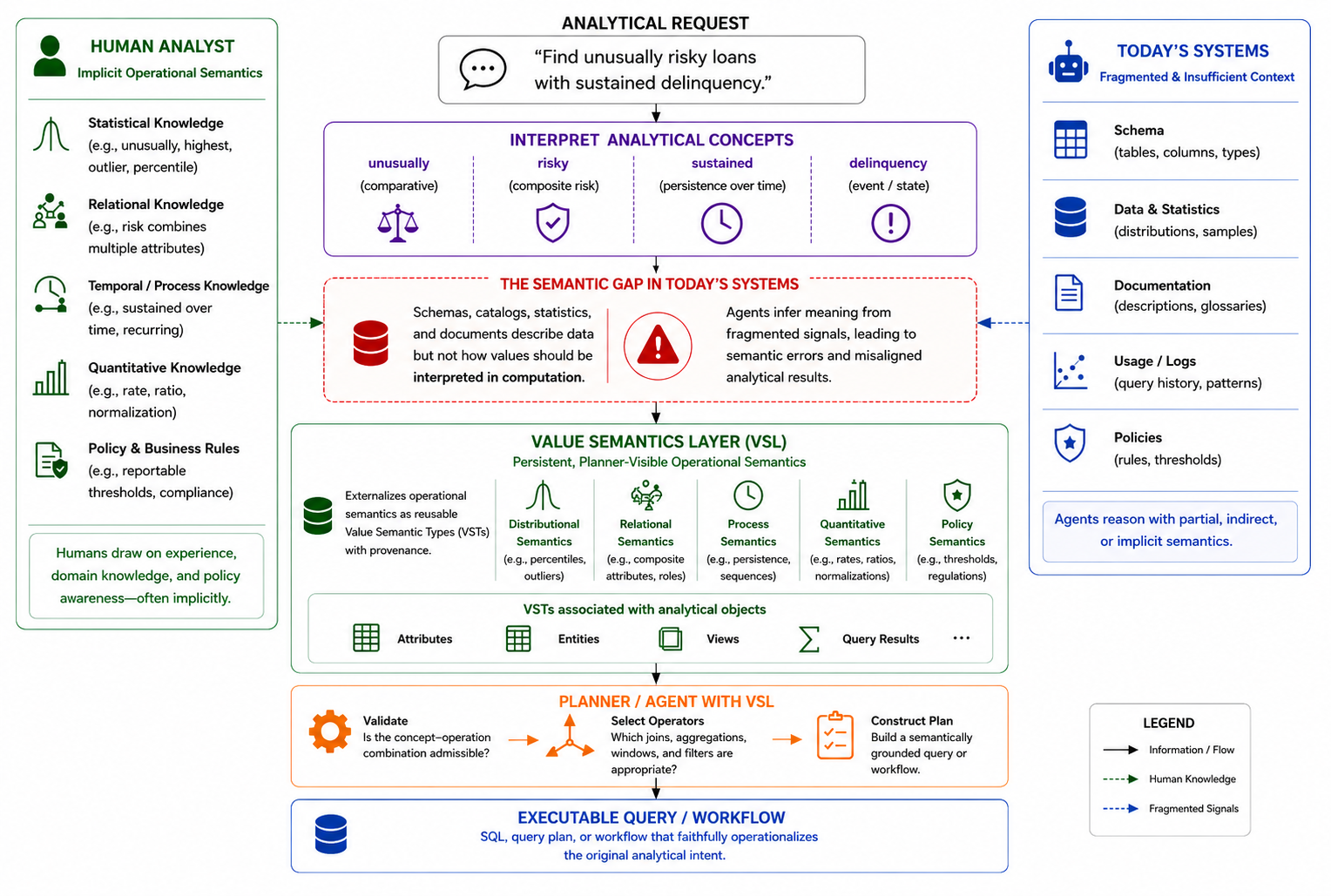}
\caption{
The Value Semantics Layer (VSL) externalizes 
value semantics as persistent planner-visible
infrastructure, enabling analytical concepts such as
\emph{unusually}, \emph{risky}, and \emph{sustained} to be
translated into executable computation.
}
\label{fig:intro_motivation}
\end{figure*}

Figure~\ref{fig:intro_motivation} illustrates this
challenge. For the request \emph{``Find unusually risky
loans with sustained delinquency,''} interpreting
\emph{unusually}, \emph{risky}, and \emph{sustained}
requires distributional, relational, process, and policy
semantics that determine 
how analytical concepts are operationalized into executable analytical workflows.
We refer to this knowledge as \emph{value
semantics}: semantic knowledge governing how values participate
in analytical operations such as operator selection, aggregation,
comparison, temporal reasoning, and policy validation.


Existing metadata systems capture schemas, lineage,
business definitions, and governance, but
provide limited support for operational semantics.\footnote{We
use \emph{operational semantics} to mean which analytical
operations are admissible for a value and under what
constraints---distinct from its programming-languages sense,
where the term denotes the meaning of a program defined by how
it executes.}
Consequently, NL2SQL
systems~\cite{wang2020rat,hong2024survey} and analytical
planners~\cite{zeighami2025proactive,Kandogan2025Blue}
rely on fragmented signals such as value statistics,
schema structure, and retrieved documentation, often
producing executable queries that nevertheless
misinterpret analytical intent. Recent analyses of agent-generated analytical
workflows~\cite{mahmud2026semanticgap} identify recurring
operationalization failures involving comparative,
process, quantitative, role, and policy semantics across
multiple domains, suggesting that many analytical concepts
require explicit operational semantics beyond schemas and
structural metadata.

We propose the \emph{Value Semantics Layer} (VSL), an
architectural abstraction that externalizes value semantics as
persistent, planner-visible infrastructure. VSL introduces
\emph{Value Semantic Types} (VSTs): reusable semantic assertions,
inferred once rather than recreated per query, that guide operator
selection, validation, and workflow generation across planners and
systems.

This paper makes three contributions:
(1) a formalization of \emph{Value Semantic Types} (VSTs)
as persistent operational semantic assertions;
(2) the \emph{Value Semantics Layer} (VSL), an
architectural abstraction for AI-native data systems; and
(3) two prototype realizations illustrating progressively
tighter planner integration.

\section{Related Work}

\textbf{Text-to-SQL and semantic grounding.}
Recent Text-to-SQL systems improve query generation through
schema linking, retrieval augmentation, and LLM-based
reasoning~\cite{wang2020rat,glass2025extractive,
wang2025linkalign,luo2025nl2sql}. Benchmarks such as
BIRD~\cite{li2023bird} supply manually curated \emph{evidence}
annotations carrying the semantic information needed for correct
value interpretation. Such evidence is \emph{query-associated},
authored per question rather than attached to the underlying
analytical objects, so the same facts are re-specified per query
rather than persisted as reusable, object-associated assertions.
VSL instead infers and formalizes such semantics once, as
persistent artifacts shared across queries and planners.

\textbf{Table language models and learned schema representations.}
Table language models and table representation
learning~\cite{yin2020tabert,herzig2020tapas,deng2020turl,yu2021grappa}
learn latent representations of tables, columns, and relational
context for parsing, retrieval, and question answering. These
learned representations are complementary to VSL rather than an
alternative: they can serve as evidence for inferring assertions,
while VSL externalizes operational semantics as explicit,
planner-visible artifacts a planner can retrieve, apply, and check
instead of leaving them implicit in model weights.

\textbf{Aggregation semantics and semantic layers.}
A long line of database research characterizes when aggregation is
valid. Summarizability and additivity analysis distinguishes
additive, semi-additive, and non-additive measures and proposes
metadata constraints that restrict potentially inaccurate aggregate
queries~\cite{lenz1997summarizability,horner2004additivity},
integrated into warehouses much as integrity constraints are.
Deployed systems go further: enterprise metadata
catalogs~\cite{apache_atlas,datahub,amundsen} persist schemas,
lineage, and governance metadata, while BI semantic layers
(LookML/Looker~\cite{lookml}, the dbt Semantic Layer /
MetricFlow~\cite{dbt_semantic_layer}, and Snowflake Semantic
Views~\cite{snowflake_semantic_views}), in the OLAP
tradition~\cite{kimball2013datawarehouse,inmon2005building,codd1993olap},
encode executable business definitions and operational behavior, not
merely metrics and documentation. Prior research and these deployed
systems thus already support aspects of admissible aggregation,
dimension-dependent additivity, persistent semantic definitions, and
the checking of invalid operations; persistence, admissibility
metadata, and enforcement are not, by themselves, new. Column
semantic-type detection~\cite{sherlock,sato} is a further useful
input, classifying \emph{what a value is} rather than \emph{how it
may be validly operated on}. VSL reuses such definitions and
constraints where available, while investigating evidence-backed
assertions over a broader range of analytical interpretations than
additivity alone. One such interpretation, temporal persistence,
differs from semi-additivity: semi-additivity concerns whether a
measure such as \texttt{balance} may be meaningfully summed across a
dimension, whereas temporal persistence concerns whether a condition
holds over an ordered sequence, requiring entity grouping, temporal
reasoning, and a supplied persistence definition. Existing OLAP
systems can compute such temporal expressions; VSL's role is to
record this requirement as a reusable operational constraint on the
attribute.

\textbf{Inference foundations and VSL's contribution.}
VSL builds on established sources of evidence and constraints: data
profiling and dependency discovery expose structural regularities,
keys, and correctness
constraints~\cite{fan2008conditional,chu2013discovering,abedjan2015profiling},
while measurement theory gives a foundational account of the
transformations admissible for each scale
type~\cite{stevens1946measurement}; VSL draws on both rather than
replacing them. Ontology-based systems and schema matching reconcile
heterogeneous schemas when explicit semantic models are
available~\cite{rahm2001survey,doan2012principles,aumueller2005coma,calvanese2009ontologies,bizer2009dbpedia}.
Against this backdrop, VSL's proposed contribution is architectural:
to \emph{integrate}, for analytical objects, (i)~evidence-backed
inference of operational assertions from data, schema, relationships,
workloads, and external definitions; (ii)~their persistence as
reusable assertions carrying admissible operations, application
conditions, grounding evidence, and confidence; and (iii)~their
application during data-agent planning and validation, with broader
composition and lifecycle management left as open research
challenges.

\section{Value Semantics Inference}
\label{sec:vst-inference}


\subsection{Problem Definition}
\label{sec:problem-definition}
We define \emph{value semantics inference} as the task of
inferring reusable value 
semantics by integrating
available evidence, including value samples, schema metadata,
distributional statistics, relational structure, workload
behavior, and domain knowledge.

\textbf{Value Semantic Types (VSTs).}
A \emph{Value Semantic Type (VST)} is a reusable semantic
artifact capturing the operational semantics of an analytical
object. Unlike primitive data types (e.g., \texttt{INTEGER},
\texttt{DATE}), which describe physical representation, or domain
ontologies, which capture conceptual knowledge, VSTs specify how an
object participates in computation: its semantic roles, admissible
operations and conditions, grounding evidence, and confidence. We
use \emph{type} in the sense of a semantic classification governing
which operations are admissible for its members, not a formal type
system; composition and subtyping are left to future work.

VSTs are associated with analytical objects (e.g.,
attributes, entities, derived views,  query results)
and
are inferred from recurring evidence in value
distributions, relational structure, workloads, process
behavior, and policy knowledge.

\textbf{Formal Model.}
\label{sec:formal}
Let $\mathcal{A}$ denote the universe of analytical objects
participating in analytical computation over a relational
database. Analytical objects include schema elements
(e.g., attributes and entities) as well as derived
analytical artifacts (e.g., views, query results,
aggregated measures, and other relational expressions).

Let $X \in \mathcal{A}$ denote an analytical object
with associated semantic evidence $\mathcal{E}(X)$.
The evidence associated with $X$ may include observable data
characteristics, schema metadata, relational structure, workload
behavior, process history, and policy knowledge, drawn from the
object itself or from broader relational, temporal, and workflow
context.

Let $\mathcal{O}$ denote the space of admissible analytical
operations (e.g., comparison, aggregation, ordering,
thresholding, joins).

A Value Semantic Type associated with analytical object
$X$ is defined as an operational semantic assertion:

\[
\mathrm{VST}_i(X)=
(S_i,\Phi_i,\Gamma_i,\mathcal{E}_i,\pi_i),
\]

where

- $S_i$ denotes one or more inferred operational semantic
roles;

- $\Phi_i\subseteq\mathcal O$ is the set of \emph{admissible
analytical operations} for those roles (e.g., aggregation and
temporal ordering are permitted; direct cross-account magnitude
comparison is not);

- $\Gamma_i$ is a set of \emph{application conditions}---predicates
constraining \emph{how} operations in $\Phi_i$ may be applied
(e.g., aggregation must group by account and preserve temporal
order): $\Phi_i$ states \emph{which} operations are admissible,
$\Gamma_i$ the conditions under which they remain valid;

- $\mathcal{E}_i\subseteq\mathcal{E}(X)$ denotes the
grounding evidence supporting the assertion;

- $\pi_i\in[0,1]$ denotes evidential confidence.

Because VSTs are inferred rather than declared, each is an
evidence-supported assertion rather than a deterministic fact, and
multiple compatible VSTs may be associated with the same object.

\[
\mathcal V(X)=
\{
\mathrm{VST}_1(X),
\dots,
\mathrm{VST}_k(X)
\}.
\]

\begin{table*}[t]
\centering
\small

\caption{Families of Value Semantic Types (VSTs), grounding evidence, representative operational roles, and examples. Roles are illustrative rather than members of a fixed ontology.}

\label{tab:vst-families}
\begin{tabular}{p{0.16\linewidth} p{0.22\linewidth} p{0.26\linewidth} p{0.30\linewidth}}
\toprule
\textbf{VST family} & \textbf{Grounding evidence} & \textbf{Representative roles} & \textbf{Examples} \\
\midrule

Distributional &
Value distributions, ordering, skew, tail behavior &
ComparativeAxis; OrdinalLevel &
\emph{ComparativeAxis} over \texttt{number\_of\_persons\_injured}, encoding monotonic ordering and extreme regimes to ground intents such as \emph{severe} or \emph{mass-casualty} \\

\midrule

Relational &
Participation in joins, filters, groupings, and referential integrity constraints; query workload evidence of attribute role usage.
&
Identifier; ForeignKeyReference; AggregationKey;
SegmentationDriver; EvidenceSignal;
ContextualDescriptor
&
\emph{EvidenceSignal} for \textit{required\_years\_experience}, preventing substitution with posting frequency 
\\

\midrule

Process  &
Event-sequence–grounded state abstraction across rows &
StateVariable; EventMarker; TransitionDerivedState; LifecycleBoundary &
\emph{TransitionDerivedState} for \emph{sustained negative balance} inferred from repeated negative transactions over time \\

\midrule

Quantitative &
Functional dependencies; aggregation scope; unit compatibility; and arithmetic compositional constraints &
DerivedMeasure; UnitBearingQuantity; RatioQuantity; CompositionalAggregate &
DerivedMeasure for total repayment, encoding compositional constraints and preventing double-counting \\

\midrule

Policy &
Externally defined domain or institutional rules &
RegulatoryCategory; EligibilityBoundary; ComplianceStatus; ThresholdDefinedClass &
\emph{RegulatoryCategory} role for \emph{moving violation}, requiring policy-defined mapping \\

\bottomrule
\end{tabular}
\end{table*}

\textbf{System Representation.}
VSTs are materialized as structured semantic
artifacts capturing semantic roles, admissible operations,
grounding evidence, and operational constraints.
Figure~\ref{fig:vst-examples} illustrates representative VSTs inferred from
observable evidence.

\begin{figure}[t]
\centering
\small

\newcommand{\vstfield}[2]{%
\textbf{#1:}\\[-0.15em]
{\raggedright #2\par}\vspace{0.45em}
}

\begin{minipage}[t]{0.48\linewidth}
\vspace{0pt}
\fbox{%
\begin{minipage}[t]{0.90\linewidth}
\vspace{0.4em}
\textbf{Distributional VST}\par\vspace{0.5em}

\vstfield{Analytical object}
{\texttt{collisions.}\\\texttt{number\_of\_persons\_injured}}

\vstfield{Primary semantic role ($S$)}
{\emph{ComparativeAxis}}

\vstfield{Admissible operations ($\Phi$)}
{ordering; ranking; thresholding}

\vstfield{Application conditions ($\Gamma$)}
{non-negative values; comparisons should preserve monotonic ordering}

\vstfield{Grounding evidence ($\mathcal{E}$)}
{heavy-tailed distribution; observed extreme-value regimes}

\textbf{Confidence ($\pi$):} 0.87
\vspace{0.4em}
\end{minipage}
}
\end{minipage}
\hfill
\begin{minipage}[t]{0.48\linewidth}
\vspace{0pt}
\fbox{%
\begin{minipage}[t]{0.90\linewidth}
\vspace{0.4em}
\textbf{Process-Oriented VST}\par\vspace{0.5em}

\vstfield{Analytical object}
{\texttt{transactions.balance}}

\vstfield{Primary semantic role ($S$)}
{\emph{TransitionDerivedState}}

\vstfield{Admissible operations ($\Phi$)}
{temporal aggregation; windowed reasoning; persistence detection}

\vstfield{Application conditions ($\Gamma$)}
{requires temporally ordered events and grouping by
\texttt{account\_id}; invalid under row-local thresholding}

\vstfield{Grounding evidence ($\mathcal{E}$)}
{repeated negative events; temporal ordering; account-level grouping}

\textbf{Confidence ($\pi$):} 0.81
\vspace{0.4em}
\end{minipage}
}
\end{minipage}

\caption{Illustrative instances of the VST formalism
$(S,\Phi,\Gamma,\mathcal{E},\pi)$ for attribute-level analytical
objects. The formalism generalizes to entities, derived views,
and other analytical objects.}

\label{fig:vst-examples}
\end{figure}

\begin{figure}[t]
    \centering
    \includegraphics[width=1.0\columnwidth]{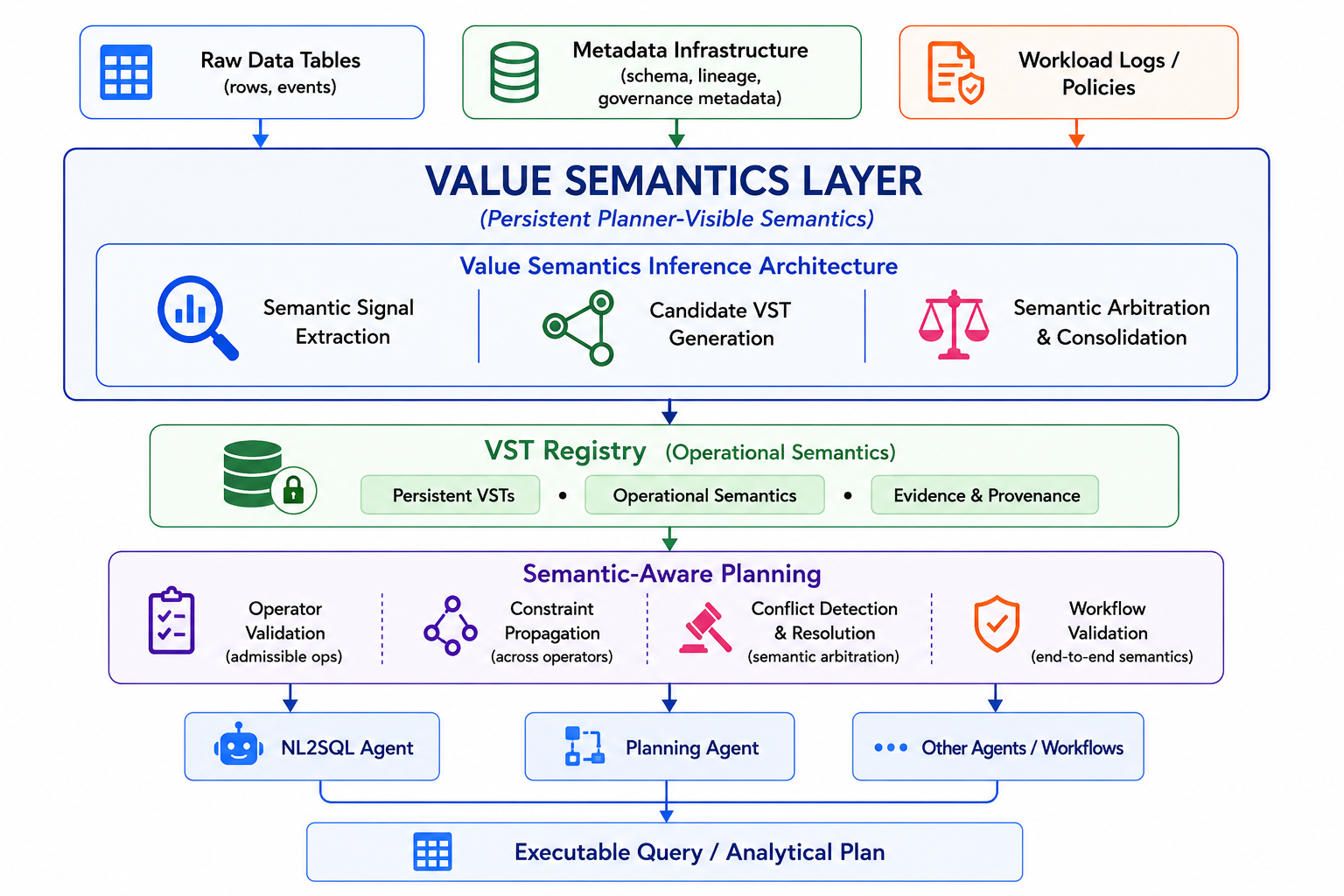}

\caption{Proposed architecture of the Value Semantics Layer (VSL).}

    \label{fig:vsl-architecture}
\end{figure}



\textbf{Recurring Families of VSTs.}
Although VSTs do not form a closed type system, inferred
operational semantics cluster into recurring families
(Table~\ref{tab:vst-families}) defined by shared grounding
patterns. These form an evolving taxonomy rather than fixed
ontology classes, expected to grow as semantic evidence expands.

\subsection{The Value Semantics Inference Architecture}
\label{sec:arch}

Value semantics leave observable traces in data
distributions, relational structure, workload behavior, and
external policies. VSL discovers reusable Value Semantic
Types (VSTs) from these heterogeneous sources of semantic
evidence through a multi-stage architecture illustrated in
Figure~\ref{fig:vsl-architecture}.

\textbf{Stage 1: Semantic signal extraction.}
\emph{Inputs:} observed values, schema metadata, relational
structure, workload behavior, process histories, and external
policy or domain knowledge.
\emph{Method:} distributional profiling, key/join and dependency
analysis, workload mining, and process-pattern detection over
ordered per-entity sequences.
\emph{Output:} \emph{semantic signatures} for candidate analytical
objects. For example, an \texttt{amount} field that co-occurs with
\texttt{interest\_rate}, participates in loan joins, and is
non-negative and right-skewed is constrained toward a financial
measure rather than a generic number.

\textbf{Stage 2: Candidate VST generation.}
\emph{Inputs:} semantic signatures.
\emph{Method:} map each signature to candidate VSTs by its
operational signature---high uniqueness with stable join
participation indicates identifier semantics (aggregation
forbidden); heavy-tailed numeric distributions indicate
comparative semantics; repeated per-entity events indicate
process-temporal semantics; arithmetic relationships indicate
quantitative roles---attaching a confidence that aggregates the
strength and agreement of the supporting signals.
\emph{Output:} a sparse set of candidate VSTs, each with a
semantic role, admissible operations $\Phi$, application
conditions $\Gamma$, grounding evidence, and confidence.

\textbf{Stage 3: Arbitration and consolidation.}
\emph{Inputs:} the candidate VSTs for an object, which may compete
on overlapping evidence.
\emph{Method:} reconcile competing candidates \emph{without}
forcing a silent guess. Candidate assertions conflict when they prescribe contradictory
validity judgments for the same operation under the same context. A candidate rule groups
conflicting assertions, selects among them by evidential support
and confidence margin, and \emph{retains low-margin conflicts as
ranked alternatives}, so a planner can make a confidence-aware
choice or request clarification. Making this arbitration reliable,
and calibrating its confidences, is an open problem
(\S\ref{sec:agenda}, C1); we present the rule as a design, not a
settled mechanism.
\emph{Output:} consolidated VST assertions, materialized as
persistent artifacts with their operations, conditions, grounding
evidence, provenance, and confidence.

\textbf{A worked example (the inference pipeline).}
Consider \texttt{balance} in the finance schema.
\emph{Stage~1 (signals):} profiling finds it numeric and signed,
dependency analysis groups it by \texttt{account\_id}, and
process-pattern detection over the per-account transaction order
finds runs of consecutive negatives---so the signature carries both
a plain-numeric and an ordered-per-entity signal.
\emph{Stage~2 (candidates):} these map to two competing readings. A
\emph{quantitative-measure} reading treats \texttt{balance} as a
magnitude admitting row-local comparison and aggregation; a
\emph{process-temporal} (\emph{TransitionDerivedState}) reading
treats it as an ordered per-account series with $\Phi=\{$aggregate,
order-by-time, window$\}$ under $\Gamma=\{$partition by account,
respect transaction order$\}$. Confidence aggregates signal strength
and agreement: the process-temporal candidate rests on three
corroborating signals, the scalar reading on one.
\emph{Stage~3 (arbitration):} under a sustained-negative-balance
intent the readings conflict---the scalar one admits a row-local
filter the process-temporal one forbids---and arbitration selects
the process-temporal reading on evidential margin; were the margin
small, it would \emph{retain both as ranked alternatives} for the
planner. The consolidated VST is persisted with its operations,
conditions, evidence, and (heuristic) confidence. A planner that
later retrieves it enforces the semantics: under a \emph{``sustained
negative balance''} intent, where $\Gamma$ applies, it rejects
\texttt{WHERE balance < 0} in favor of a windowed formulation over
the ordered sequence. Inference fixes the \emph{role} but not the
persistence criterion: what counts as ``sustained'' is a
definitional choice from a business rule or clarification, not from
the value distribution alone.

Policy-grounded VSTs additionally encode mappings derived from
external regulations, institutional rules, or domain definitions;
these often depend on multiple objects or workflow-level conditions
that cannot be recovered from observed data, letting planners reason
with externally defined semantics.

\textbf{Realization and feasibility.}
Our implementation infers VSTs at the attribute level; future
realizations may extend inference to entities, derived views, and
workflow fragments under the same principles. Because the
architecture builds on established database techniques (profiling,
dependency discovery, workload analysis, metadata management,
constraint validation), it is designed for incremental deployment
without modifying storage or query engines, with a registry that
\emph{would} integrate across engines through existing metadata
interfaces; our prototype realizes this for a single NL2SQL agent,
and cross-engine generalization is a design goal, not yet
demonstrated. Unlike descriptive metadata, retrieved VSTs are
consumed during planning: the planner restricts operators to $\Phi$,
checks candidate SQL against $\Gamma$, and weighs evidence
$\mathcal{E}$ and confidence $\pi$ in deciding how far to trust it.
Acting on an uncertain assertion is itself open---reject, advise, or
request clarification---and propagating these constraints
compositionally across stages is proposed, not demonstrated
(\S\ref{sec:agenda}, C2).

\section{Prototype Realization and Preliminary Evaluation}

To assess feasibility, we implemented two realizations of the
Value Semantics Layer and integrated them with the Blue NL2SQL
agent~\cite{Kandogan2025Blue}. The two realizations compare using persistent VSTs as prompt
context with integrating them into planning and validation.

\textbf{Prototype I: Prompt-mediated (PM).}
The first realization exposes VSTs through prompt-mediated
semantic grounding. VST annotations are serialized together
with schema metadata and injected into the LLM prompt,
providing semantic guidance during SQL generation without
modifying the underlying planner.

\textbf{Prototype II: Planner-enforced (PE).}
The second realization integrates VSTs into a lightweight VSL
planner backed by the registry: it retrieves intent-relevant VSTs,
injects their admissible operations $\Phi$ and conditions $\Gamma$
into the plan, and validates the generated SQL with a
deterministic checker (\texttt{sqlglot} parse plus VST rules). The
checker flags a fixed set of violations: aggregation,
\texttt{SUM}/\texttt{AVG}, magnitude comparison, or ordering
applied to an identifier; aggregations a VST forbids; and
row-local filtering of a process-temporal attribute that requires
temporal aggregation or windowing. Each flagged violation triggers
targeted regeneration (the \emph{rejection loop}).

\textbf{Implementation scope.}
Signal extraction, rule-based candidate generation with heuristic
confidence, a confidence-thresholded consolidator, and the
deterministic validator are implemented. The
registries used in the evaluation were generated using the
inference engine. The evaluation uses the resulting registries in both
VSL realizations. Three mechanisms remain
\emph{unimplemented}, together constituting the open problem of
reliable inference and maintenance (\S\ref{sec:agenda}, C1):
calibrating assertion confidence against empirical reliability;
surfacing low-margin conflicts as ranked alternatives and
requesting clarification rather than thresholding to a single
interpretation; and revalidating assertions as data, workload, and
policy change.

\textbf{Evaluation.}
Both realizations were evaluated on a curated diagnostic subset of
the formative study in~\cite{mahmud2026semanticgap}. From its
236-intent collection, we selected 124 cases labeled as non-policy
operationalization failures, excluding 29 policy-grounding cases and
83 cases labeled as having no operationalization failure. These
cases span Finance, Public Safety, and Human Resources across
comparative, process-temporal, quantitative, and role-oriented
semantics.
VSL's semantic categories were motivated by the taxonomy
in~\cite{mahmud2026semanticgap}. The PE prototype uses
domain-specific intent patterns, some of which overlap with the
evaluated query phrasings; its generalization to unseen
formulations remains untested. The prompt-mediated (PM) condition
does not use these planner rules.
These categories instantiate the VST families of
Table~\ref{tab:vst-families} (comparative and role-oriented map to
the \emph{Distributional} and \emph{Relational} families;
process-temporal and quantitative to \emph{Process} and
\emph{Quantitative}); the \emph{Policy} family lies outside this
evaluation, whose 124 cases are all non-policy.
Schemas, models, and execution settings are identical across all
conditions; only the VSL integration differs. All used GPT-4o with
identical decoding and a single run per case, as in the formative
study~\cite{mahmud2026semanticgap}.
Operationalizations are classified as fully resolved,
partially resolved, or unresolved according to whether the
intended analytical concept is correctly operationalized.
Resolution labels were assigned by the authors.
We also evaluate two non-VSL controls on the same 124 cases: a
\emph{data-dictionary} baseline (schema plus GPT-4o-generated,
author-reviewed plain column descriptions, operational semantics
deliberately excluded) and an \emph{information-matched} prose
control (the same value-semantic facts the VSTs encode, as prose
without their structure or enforcement). The controls separate
VSL's structured, enforceable representation from the information a
catalog already exposes.

\begin{table}[t]
  \centering
  \small
  \caption{Semantic grounding on the 124
  operationalization failures from~\cite{mahmud2026semanticgap}, by
  semantic category. PM = Prompt-mediated VSL; PE = Planner-enforced
  VSL; Fully/Partially = fully/partially resolved;
  Unresolved $= n -$ Fully $-$ Partially. The data-dictionary and
  matched-prose controls are defined in the text; the final row
  reports fully-resolved totals for all four conditions.}
  
  \label{tab:vsl_results}
  
  \begin{tabular}{lcccccc}
  \toprule
  & & \multicolumn{2}{c}{\textbf{PM}} &
  \multicolumn{2}{c}{\textbf{PE}} \\
  \cmidrule(lr){3-4}\cmidrule(lr){5-6}
  \textbf{Category} &
  \textbf{n} &
  \textbf{Fully} &
  \textbf{Partially} &
  \textbf{Fully} &
  \textbf{Partially} \\
  \midrule
  Comparative  & 43 & 18 & 23 & 24 & 18 \\
  Process       & 40 &  3 & 25 & 16 & 17 \\
  Quantitative  & 20 & 11 &  8 &  8 & 12 \\
  Role          & 21 &  3 & 17 &  3 & 17 \\
  \midrule
  Total         &124 & 35 & 73 & 51 & 64 \\
  \midrule
  \multicolumn{6}{l}{\footnotesize Fully resolved of 124:
  data-dictionary 17, matched prose 20, PM 35, PE 51.}\\
  \bottomrule
  \end{tabular}
  \end{table}

\textbf{Results.}
Table~\ref{tab:vsl_results} summarizes the results. We treat
\emph{fully resolved} operationalizations---where the intended
analytical concept is correctly operationalized---as the primary
measure, since partial credit (overall grounding) is far less
discriminating. Planner-enforced VSL raises fully resolved
operationalizations from 35 to 51 (overall grounding rises from
87.1\% to 92.7\%), providing preliminary evidence of a benefit of
planner-visible operational semantics over prompt-mediated guidance. PE supplies the applicable operations and conditions before SQL
generation and triggers regeneration when the validator flags a
violation. Regeneration was triggered in $6\%$ of cases. The
evaluation does not separately estimate the effects of planning
guidance and validation-triggered regeneration.

The two controls (Table~\ref{tab:vsl_results}) resolve far fewer
cases (data-dictionary 17 and matched prose 20 of 124) than
prompt-mediated (35) or planner-enforced (51) VSTs, despite
comparable loose grounding (both controls $81.5\%$, vs.\ PM
$87.1\%$ and PE $92.7\%$). The information-matched prose control,
which carries the same value-semantic facts as the VSTs but as
prose, is the relevant comparison for separating information from
representation: its gap to prompt-mediated VSTs ($20$ to $35$)
is consistent with a benefit from structured semantic guidance,
and the further gain from PM to PE ($51$) with tighter planner
integration. The data-dictionary baseline is
description-only by construction, so its $17$ cases show that
descriptive metadata alone is insufficient, not that VSTs surpass
catalogs enriched with operational facts; neither control isolates
enforcement from the planning constraints it is bundled with.

The largest gains occur for process-temporal semantics, where
planner integration raises fully resolved cases from 3 to 16 by
enforcing persistence and temporal constraints during planning.
Comparative semantics also improve, through planner-guided
distribution-aware thresholds. Quantitative semantics benefit less,
suggesting many such concepts are already specified by persistent
VST annotations alone, while planner guidance occasionally
over-constrains valid formulations. Role semantics remain the
hardest category: the ambiguity lies in selecting the intended
analytical proxy rather than operationalizing it, so some concepts
will require interactive clarification or explicit business
definitions beyond value semantics.

Overall, these results are preliminary evidence that operational
semantics help most when elevated from prompt context to
planner-visible infrastructure. Planner enforcement nonetheless
remains limited by assertion coverage and validator capabilities:
it did not detect an unsupported categorical filter or an invalid
progression comparison in one evaluated query. This evaluation
targets a curated set of operationalization failures across three
domains; validation on standard Text-to-SQL benchmarks is an
explicit near-term milestone.

\textbf{Overhead.}
The local machinery VSL adds is negligible beside language-model
generation: loading a registry ($23$--$71$ VSTs) costs under
$2$\,ms once, and per-query planning plus validation about $1$\,ms
together (medians $0.37$ and $0.77$\,ms), against generation
latencies of seconds. These are local costs, not end-to-end latency:
the rejection loop adds model calls only on a detected violation
($6\%$ of the 124 queries, $0.09$ extra on average). Retrieval
scales with registry size; indexing at scale is future work.

\section{Research Agenda and Open Challenges}
\label{sec:agenda}

Realizing VSL is a systems problem, not a prompting one. We frame
the open challenges as three.

\textbf{C1 --- Reliable inference, representation, and maintenance.}
A single object may admit several competing VST hypotheses from
different signals (distributional, relational, workload, process).
The challenge is to map signals to candidates, \emph{calibrate}
confidence, and \emph{arbitrate} conflicts by evidential support and
compatibility, not defer them to consumers; and, as registries grow,
to \emph{represent} and index VSTs for retrieval at scale and
\emph{maintain} them across their lifecycle (ownership, human review
and override) as data, workloads, and policies change. Success means
calibrated confidence,
arbitration evaluated on a labeled multi-hypothesis benchmark,
sub-linear retrieval at scale, and drift-triggered revalidation.

\textbf{C2 --- Compositional semantic validation.}
The current realization validates attribute-level operations. The
harder problem is propagating value semantics through
\emph{compositions}---derived expressions, intermediate results,
and multi-step plans---where operations combine attributes of
different semantics and a locally valid step can be globally
invalid. For example, a join on a non-unique key can duplicate a
measure before aggregation: each step executes, yet the composed
result double-counts, though neither step is invalid alone. This
requires rules for how admissible operations and conditions compose,
extending validation from single operations to whole plans and
tool-using agents. Success means validation demonstrated for derived
objects and multi-step plans.

\textbf{C3 --- External grounding and interoperability.}
Some semantics cannot be inferred from data: policy-grounded
notions (e.g., ``high-risk,'' ``moving violation'') must be
acquired from authoritative sources and kept current, and existing
semantic assets---BI metric definitions, catalogs, ontologies---%
should be ingested rather than re-derived. Realizing VSL as \emph{shared} infrastructure further requires a
portable interchange format and aligned VST vocabularies across
organizations. Success means a provenance-linked acquisition
pipeline and cross-system alignment.

\section{Conclusion}

We have argued for the Value Semantics Layer (VSL): databases that
expose value-semantic assertions alongside schemas, constraints, and
statistics, letting agents retrieve and check the conditions
governing their plans rather than reconstruct that knowledge per
query. Realizing it requires reliable inference and maintenance,
compositional validation, and interoperable grounding---the
challenges set out above.
\bibliographystyle{ACM-Reference-Format}
\bibliography{references}

\end{document}